# · Giant interfacial Dzyaloshinskii-Moriya Interaction of epitaxial perovskite $La_{0.7}Sr_{0.3}MnO_3$ films

Liu Yang,[1,2][*] Xiaotong Zhang,[3][*] Hanchen Wang,[1,2][*] Na Lei,[1,2][†] Jinlong Wang, [1,2] Yiming Sun,[1,2] Lei Liu,[3] Zhiyuan Zhao,[4] Yumin Yang,[4] Dahai Wei, [4] Dong Pan,[4] Jianhua Zhao,[4] Jian Shen, [3,5,6] Weisheng Zhao, [1,2] Haichang Lu,[1,2][†] Wenbin Wang,[3,5,6][†] Haiming Yu[1,2][†]

[1] Fert Beijing Institute, MIIT Key Laboratory of Spintronics, School of Integrated Circuit Science and Engineering, Beihang University, Beijing 100191, China

[2]National Key Lab of Spintronics, Institute of International Innovation, Beihang University, Yuhang District, Hangzhou, 311115, China

[3] State Key Laboratory of Surface Physics, Institute for Nanoelectronic Devices and Quantum Computing, and Department of Physics, Fudan University, Shanghai 200433, China

[4]State Key Laboratory for Superlattices and Microstructures, Institute of Semiconductors, Chinese Academy of Sciences, Beijing 100083, China

[5]Shanghai Research Center for Quantum Sciences, Shanghai 201315, China

[6] Zhangjiang Fudan International Innovation Center, Fudan University, Shanghai 201210, China

[*]These authors contributed equally to this work.
[†]Corresponding authors: na.lei@buaa.edu.cn ; HaichangLu@buaa.edu.cn ; wangwb@fudan.edu.cn; haiming.yu@buaa.edu.cn

ABSTRACT

The Dzyaloshinskii-Moriya interaction (DMI) plays a critical role in stabilizing topological spin textures, a key area of growing interest in oxide-based spintronics. While most of reported topological phenomena found in manganites are related to the bulk-like DMI, the understanding of interfacial DMI and its origin in oxide interfaces remain limited. Here we experimentally investigate the interfacial DMI of $La_{0.7}Sr_{0.3}MnO_3$ (LSMO) films grown on various substrates by employing spin-wave propagation with drift velocities at room temperature. Our findings reveal a giant interfacial DMI coefficient ($D_s$) of 1.96 pJ/m in $LSMO/NdGaO_3$(110) system, exceeding previously reported values in oxides by one to two orders of magnitude. First-principles calculations further show that with the aid of 6$s$ electrons, the 4$f$ electrons from Nd play a key role in enhancing the spin-orbit coupling of the 3$d$ electrons in Mn, ultimately leading to the observed giant interfacial DMI. This discovery of giant interfacial DMI through engineering the interface of oxides provides valuable insights for advancing functional chiral magnonics and spintronics.

## INTRODUCTION

Chiral spin textures, such as magnetic skyrmions and chiral domain walls (DWs), hold great promise for chiral spintronics due to their potential to enable enhanced performance and novel functionalities [1,2]. These textures can be stabilized by a specific energy term that favors tilted spins between magnetic moments, known as the Dzyaloshinskii-Moriya interaction (DMI). The DMI amplitude is crucial in attaining efficient DW motion [3,4], deterministic field-free and complete magnetization switching [5,6], and high-density spintronic memories [7,8]. Therefore, controlling the strength and sign of DMI is of paramount importance.

DMI originates from a combination of strong spin-orbit coupling and broken inversion symmetry [9,10,11]. It manifests in both bulk and interfacial forms [12,13,14], where the interfacial environments offer a unique advantage: the ability to tailor the sign and strength of DMI. This has been extensively explored in metallic ferromagnets (FMs) [15,16,17]. Compared to conventional metallic FM multilayers, interfaces in magnetic oxides present a promising platform for fine-tuning magnetic chirality through charge transfer, oxygen octahedral distortion, proximity effects, and so on [18, 19]. Additionally, magnetic oxides exhibit several inherent advantages as a platform for chiral spin textures, including ultra-low magnetic damping [20], large magnon diffusion length [21], and fast DW motion [22,23]. The chiral spin textures have been reported in various oxide heterostructures including LSMO/SrRuO$_3$ [24,25], LSMO/SrIrO$_3$ [26,27], Y$_3$Fe$_5$O$_{12}$ (YIG) and Tm$_3$Fe$_5$O$_{12}$ (TmIG) on Gd$_3$Ga$_5$O$_{12}$ (GGG) [28,29]. However, the experimental estimation of interfacial DMI coefficients in Y$_3$Fe$_5$O$_{12}$ (YIG) and Tm$_3$Fe$_5$O$_{12}$ (TmIG) on Gd$_3$Ga$_5$O$_{12}$ (GGG) all fall below 0.13 pJ/m, which are significantly weaker than those found in Pt/Co-based metallic multilayers (by two orders of magnitude). This weak interfacial DMI hinders potential applications and necessitates further exploration of its physical origin in oxides. Interestingly, recent DFT calculations predict a much stronger DMI at the $La_{2/3}Ba_{1/3}MnO_3/SrTiO_3$ perovskite oxides interface, reaching 3.84 mJ/m² at 10 K [30]. However, experimental verification of significant interfacial DMI remains elusive.

In this work, we explore the interfacial DMI in $La_{0.7}Sr_{0.3}MnO_3$ (LSMO) thin films and investigate the impact of varying substrate-induced orbital hybridization by depositing LSMO on $(LaAlO_3)_{0.3}(Sr_2AlTaO_6)_{0.7}$ (LSAT), $SrTiO_3$ (STO), and $NdGaO_3$ (NGO) substrates. Chiral spin-wave velocities are measured to quantify the interfacial DMI. Our key finding is an exceptionally large interfacial DMI coefficient ($D_s$) of 1.96 pJ/m in the LSMO/NdGaO₃ (110) system. This result surpasses previous experimental values in oxide materials by one to two orders of magnitude. We further demonstrate a proportionality between $D_s$ and interfacial magnetic anisotropies ($K_i$), suggesting a common origin in interfacial spin-orbit coupling, which is confirmed by our first principal calculations. These findings pave the way for manipulating interfaces in perovskite oxides for future advancement in oxide-based chiral spintronics.

2. EXPERIMENT

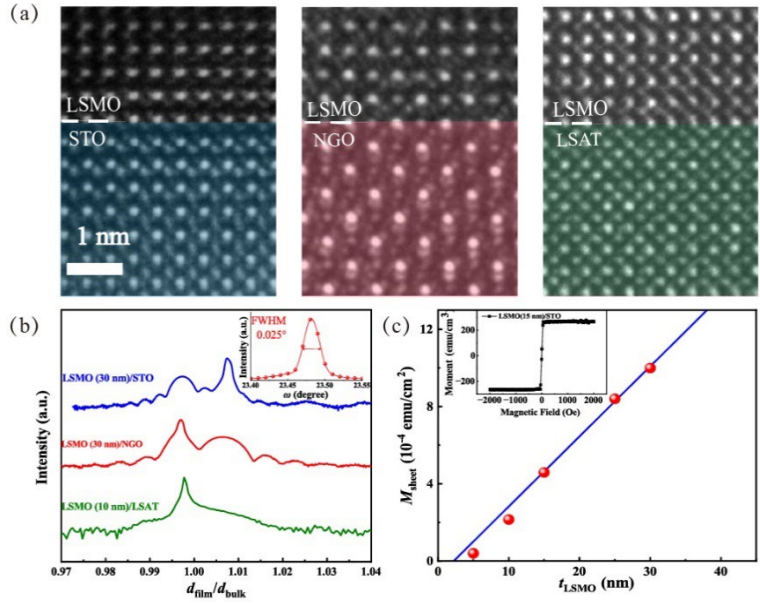

FIG. 1. (a) Transmission electron microscopy (TEM) images of LSMO films deposited on various substrates: STO, NGO and LSAT. The interfaces between the LSMO films and the substrate are marked. (b) X-ray diffraction (XRD) spectra of 10 nm LSMO/LSAT (001), 30 nm LSMO/STO (001), 30 nm LSMO/NGO (110). The spectra are offset for clarity. $d_{bulk}$ and $d_{film}$ is the interplanar (002) distance of the bulk LSMO and LSMO films on different substrates. The inset is $\omega$-scan of 30 nm LSMO/NGO (110), with a full width at half maximum (FWHM) of 0.025°. (c) Magnetic moment per unit area ($M_{sheet}$) of the LSMO film as a function of its thickness, which are measured at 300 K. The blue line represents a linear fit. The inset displays the magnetization versus magnetic field (*M-H*) loops for 15 nm LSMO/STO (001), measured at 300 K.

$La_{0.7}Sr_{0.3}MnO_3$(001) films with thickness ranging from 5 to 35 nm, were epitaxially grown on the STO (001), LSAT (001) and NGO (110) substrates using pulsed laser deposition. The KrF excimer laser was used with a pulse repetition rate of 2 Hz and a energy density of 2 J/cm$^2$. During growth, the substrate was kept at 750°C under an oxygen pressure of 1 mbar with 10% ozone. The single-crystal $SrTiO_3$(001) substrates were etched using a buffered hydrofluoric acid solution followed by oxygen annealing at 950 °C to achieve $TiO_2$-termination. The growth process is *in-situ* monitored by reflection high-energy electron diffraction (RHEED), whose intensity oscillation spectrum is depicted in Fig. S1, indicating the layer-by-layer growth mode of LSMO films (For more complete details, please refer to Supplementary Materials S1). The transmission electron microscopy (TEM) images of the LSMO layers on various substrates are presented in Fig. 1(a), revealing epitaxial films with well-defined interfaces. Moreover, the high-resolution X-ray diffraction (XRD) is performed to investigate the full LSMO structures, and the clean LSMO (002) diffraction peaks are observed, as shown in Fig. 1(b). The $d_{film}/d_{bulk}$ parameter illustrates the ratio of the interplanar (002) distance of bulk LSMO and LSMO films deposited on different substrates which enable a comparison of the strain state of films. The $d_{films}/d_{bulk}$ of STO is beneath 1, suggesting the presence of in-plane tensile strain induced by STO. Compared to STO, LSAT and NGO bring the in-plane compressive strain to the LSMO (the $d_{films}/d_{bulk}$ ratio > 1). In addition, NGO imposes a larger strain than LSAT. The inset of Fig. 1(b) shows the corresponding rocking curve spectrum of the LSMO (002) peak.

The peak has a full width at half maximum (FWHM) of only 0.025°, indicating the high crystalline quality of the 30 nm on NGO (110). Magnetic properties are characterized using a Quantum Design MPMS-3 SQUID magnetometer at room temperature. Both the in-plane and out-of-plane hysteresis loops are shown in Fig. S2 (see Supplementary Materials S2 for details). Figure 1(c) plots the magnetic moments per unit area of LSMO films on STO (001) as a function of film thickness measured at 300 K. A linear relationship is observed, with a critical nonmagnetic layer thickness of ~ 2.3 nm. Note that the Curie temperature of 5 nm LSMO is around room temperature [31]. The inset of the Fig. 1(c) shows the *M-H* loops of 15 nm LSMO film on STO measured at 300 K. The saturation magnetic moment per Mn atom is approximately 2.1 $\mu_B$/Mn.

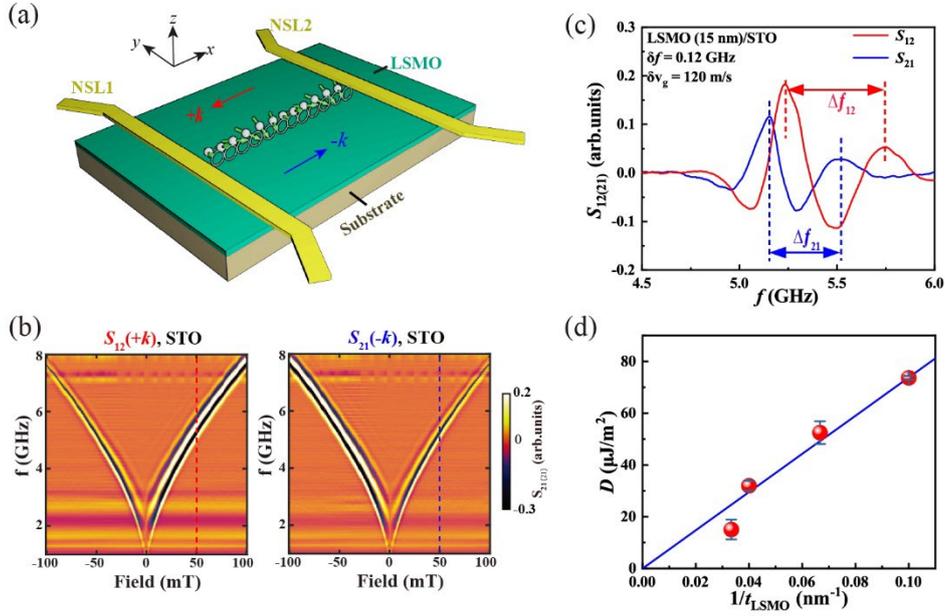

FIG. 2. (a) Schematic of the spin-wave propagation measurement setup. Two identical nano-straplines (NSLs) are used for both exciting and detecting the spin waves. (b) Transmission spectra ($S_{12}$ and $S_{21}$) of 15 nm LSMO/STO (001) obtained at DE configuration, with the magnetic field sweeping from -100 mT to +100 mT. $S_{12}$ ($S_{21}$) presents the spin wave excited by NSL2 (NSL1) and detected by NSL1 (NSL2). (c) The extracted line profiles from the spectra at an external magnetic field of +50 mT in Fig. 2(b). The corresponding DMI-induced spin-wave group velocity difference $\delta v_g$ extracted from the guided blue and red dashed lines is about 120 m/s. (d) The thickness dependence of the Dzyaloshinskii-Moriya interaction constant in LSMO/STO (001) system. A linear

fit is shown by the blue line, indicating the interfacial origin of the observed DMI.

The DMI of LSMO films on different substrates is characterized by measuring the DMI-induced drift spin-wave group velocity, which is chiral with respect to the external magnetic field and normal vector [29,32,33]. To achieve this characterization, two nano-straplines (NSLs) made of Ti (10)/Au (100) separated by 1 μm are fabricated on the LSMO surface using electron-beam lithography followed by magnetron sputtering and lift-off. Figure 2(a) shows a schematic of the measurement setup. Two NSLs are used to both excite and detect the propagating spin waves with chiral group velocities in LSMO thin films [34, 35]. The spin wave propagation direction (represented by wavevector $k$) is orthogonal to the applied in-plane external magnetic field ($H$). Figure 2(b) shows the spectra of the Damon-Eshbach (DE) spin-wave transmission spectra in opposite directions, denoted as $S_{12}(+k)$ and $S_{21}(-k)$. These spectra exhibit a noticeable asymmetry, indicating a distinct difference between spin waves propagating in opposite directions. The spin-wave group velocity for each direction can be extracted from the spectra of LSMO (15 nm)/STO by following $v_g = \frac{dw}{dk} = \frac{2\pi f}{2\pi/s} = s \cdot \Delta f_{12(21)}$, where the spin-wave propagation distance $s = 1$ μm. According to $\delta v_g = v_{g_{12}} - v_{g_{21}}$, where $v_{g_{12}} = 490$ m/s, and $v_{g_{21}} = 370$ m/s, we can get $\delta v_g = 120$ m/s in Fig. 2(c). Following the relation derived in Ref. [36],

$$\delta v_g = 2 * v_g^{DMI}, \qquad (1)$$

$$v_g^{DMI} = [(\hat{n} \times \hat{H}) \cdot \hat{k}] \frac{2\mu_0 \gamma}{M_s} D, \qquad (2)$$

where $\hat{n}$ is the film normal vector, $\hat{H}$ is the external magentic field, $D$ is the DMI constant, $M_S$ is the saturation magnetization, and $\gamma$ is the gyromagnetic ratio. By substituting the values ($M_S = 306$ kA/m for the 15 nm LSMO film and $\mu_0 \hat{H} = 50$ mT) into Eq. (2) and the relationship from Eq. (1), we can deduce a DMI constant of $D = 52.5$ μJ/m². At positive fields, the $S_{12}$ is faster than $S_{21}$ which indicates that the sign of DMI is negative [29]. To determine the origin of the DMI, we measured the DMI of LSMO films with varying thicknesses on STO (001). The DMI constants are plotted in Fig. 2(d) and exhibit a clear linear relationship with $1/t$ that intercepts the interfacial origin. This indicates that the DMI arises from contributions at the interface. To verify the contribution of the Rashba effect to the interfacial DMI, we prepared the SrO-terminated STO (001) substrate with a reversed polar direction compared to $TiO_2$-terminated case [37]. This substrate is obtained through high-temperature annealing in air (1300°C for three days) [38,39]. The typical surface morphology is confirmed by AFM image as shown in Fig. S3(a). Transmission spectra with $\pm k$ at 50 mT are presented in Fig. S3(b). The estimated $D$ of ~38 $\mu J/m^2$ is comparable to the value obtained for the $TiO_2$-terminated STO substate (see Supplementary Materials S3 for details). These results suggest that the Rashba effect at the LSMO/STO interface plays a minor role in the interfacial DMI and can be neglected.

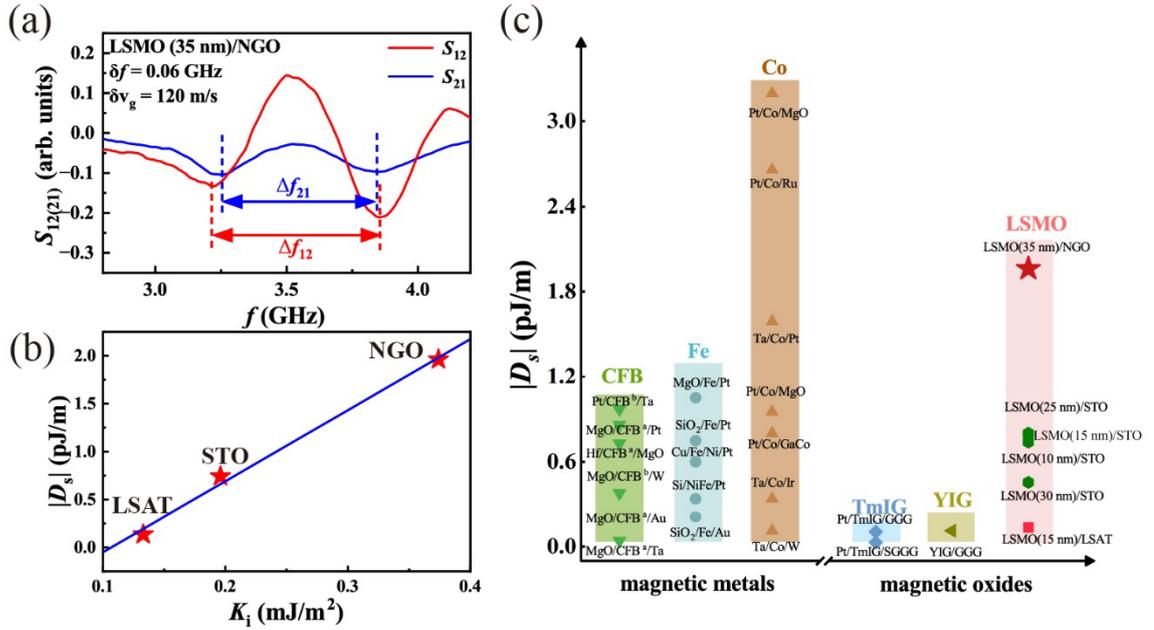

FIG. 3. (a) Transmission spectra ($S_{12}$ and $S_{21}$) of LSMO (35 nm)/NGO (110) at an applied magnetic field of 50 mT. (b) Dependence of the DMI strength on interfacial magnetic anisotropy ($K_i$) of LSMO deposited on three substrates. (c) Summary of interfacial Dzyaloshinskii-Moriya interaction in various magnetic metals and oxides [24]. Here, CFB[a] refers to $Co_{20}Fe_{60}B_{20}$, and CFB[b] refers to $Co_{40}Fe_{40}B_{20}$.

To further investigate the presence of interfacial DMI at oxide interfaces and elucidate its microscopic origin, we measured DMI on various substrates. We observed a large interfacial DMI value at the interface between LSMO and NGO. As shown in Fig. 3(a), the difference in group velocities reaches 120 m/s for 35 nm LSMO on NGO substrate (with a spin-wave propagation distance of 2 μm on NGO). By substituting these values into equations (1) and (2), along with $M_S$ of 333 kA/m, we estimate a DMI constant of 56 μJ/m$^2$ for 35 nm LSMO/NGO (110). In contrast, the DMI for LSMO on LSAT appears to be significantly weaker, with a value of only 9 μJ/m$^2$ (see Supplementary Materials S3 for details). To facilitate a direct comparison between different systems, the film thickness was normalized to obtain an interfacial DMI constant ($D_s$), defined as $D_s = D \cdot t$, where $t$ represents the thickness of the LSMO layers. This normalization allows for a lateral comparison of $D_s$ across various systems. The LSMO interfacial DMI constants for LSAT, STO and NGO substrates are 0.135, 0.74, and 1.96 pJ/m, respectively. These results demonstrate that the interface can effectively regulate DMI. The interfacial DMI and the difference of the surface anisotropies at the two film surfaces both contribute to frequency nonreciprocity, but they have a very distinct dependence on film thickness. The frequency nonreciprocity induced by interfacial DMI is proportional to $t^{-1}$, while the frequency nonreciprocity induced by the difference in surface anisotropies is proportional to $t^2$ [40]. Since the experimentally measured $\delta f$, plotted in Fig. S5(a), exhibit a clear linear relationship with $1/t$, we can reasonably exclude the contribution from the difference in surface anisotropies. As for the inherent physical origin of interfacial DMI and surface anisotropy, we further provide supporting evidence for the strong correlation between the DMI and surface anisotropy [41] (detailed in Supplementary Materials S7).

Epitaxial strain disrupts the degeneracy of 3$d$ orbitals, favoring in-plane ($3d_{x^2-y^2}$) or perpendicular ($3d_{3z^2-r^2}$) occupancy based on the out-of-plane strain type (compressive or tensile) [42]. In LSMO/STO (001), out-of-plane compressive strain promotes $3d_{x^2-y^2}$ occupancy, while out-of-plane tensile strain in LSMO/NGO (110) favors $3d_{3z^2-r^2}$ [43]. Consequently, LSMO/NGO (110) exhibits the weakest in-plane magnetic anisotropy.

Figure 3(b) visually demonstrates the clear linear dependence of both interfacial magnetic anisotropy and DMI on the substrate. The LSMO/LSAT system displays the lowest $K_i$ value, mirroring the trend observed in interfacial DMI. Conversely, the substantial orbital symmetry breaking at the LSMO/NGO interface, originating from strong spin-orbit coupling, leads to a significant interfacial DMI. This correlation between interfacial DMI and $K_i$ suggests a common underlying mechanism for both parameters.

As shown in Fig. 3(c), we compared the amplitudes of the interfacial DMI constant in our LSMO films with those reported in the literatures for both metallic and oxide systems [29, 44, 45]. Currently, the metallic heterostructures, particularly Pt/Co-based configurations stand as the frontier with the highest reported interfacial DMI constant of 3.2 pJ/m [46]. In contrast, the upper limit reported for garnet films is only 0.1 pJ/m [14], an order of magnitude lower. Interestingly, the interfacial DMI constants of our LSMO films span a wider range, from 0.135 to 1.96 pJ/m, approaching the record held by metallic Pt/Co-based systems. Note that the DMI constant reported for $La_{0.7}Sr_{0.3}Mn_{1-x}Ru_xO_3$ multilayers (11.04 pJ/m) likely incorporates not only the interfacial contribution but also the effect of Ru substitution for Mn [30]. These results highlight the promise of LSMO, the prototypical perovskite, as a platform for stabilizing chiral spin textures [24,25].

To elucidate the interfacial interaction mechanisms in these heterostructures, we performed first-principles calculations on LSMO/STO (001) and LSMO/NGO (110) interfaces. The atomic structures of these interfaces are shown in Fig. 4. To rule out the effect of strain, the lattice parameters of LSMO on both STO and NGO are identical to that of the pristine bulk LSMO. The interfacial LSMO experiences less distortion when in contact with NGO than STO. We calculate the average spin-orbit coupling (SOC) matrix element of the $3d$ orbitals of the interfacial Mn atoms, as shown in Figs. 4(c) and 4(d). The SOC matrix of the interfacial Mn in the LSMO-STO system is close to that of bulk LSMO, while the SOC matrix in LSMO-NGO is demonstrably larger (Fig.

S6). Among all the SOC matrix elements, $<d_{z^2}|\hat{H}_{SOC}|d_{xz}>$ is the most significant contributor to the interfacial DMI $d_y=\frac{J_{zx}-J_{xz}}{2}$, where $\hat{H}_{SOC} \propto \hat{s}\cdot\hat{l}$ is the SOC operator and $J_{xz}$ is the off-diagonal element of the bilinear tensorial exchange. Figure S6(b) shows the variation of matrix element $<d_{z^2}|\hat{H}_{SOC}|d_{xz}>$ concerning the layer number of interfacial layers (1= interfacial Mn, 2=second layer Mn, etc.). Inspecting the Mn atoms from the interface to a few layers inside, the SOC energy is expected to converge to that of the bulk LSMO.

To investigate the mechanism of the enhanced SOC matrix in the LSMO-NGO system, we performed total energy calculations of the ferromagnetic ground state for this heterostructure, using two different pseudopotentials of Nd. One of them is to freeze three valence *4f* electrons in Nd. We found that the SOC matrix is smaller than using the pseudopotential that fully relaxes all the *f* electrons, resembling the values of bulk LSMO and LSMO-STO heterostructure, shown in Fig. S6(b). This reveals that *f* electrons play an important role in affecting the SOC effect of the interfacial Mn. Since *f* orbitals are local, they can't overlap with the *d* orbitals of Mn, which are about 3.5 Å apart. However, the 6*s* electrons of Nd are nonlocal and can serve as a medium to interact with both *d* and *f* orbitals. Freezing the *f* electrons into the core makes no interaction between them and the *s* electrons. Consequently, the formation of the DMI is solely by breaking the symmetry at the interface, which is the same as in LSMO-STO. Whereas the proximity effect takes charge as we fully relax the *f* electrons. Additionally, we also found the anisotropy in the *xy* plane can be negligible in the cases of bulk LSMO, LSMO-STO, and LSMO-NGO with frozen *f* electrons. However, it is not true for LSMO-NGO. As NGO is zigzag deformed from the perovskite cubic structure, the shape of *f* orbitals affects the in-plane isotropy of Mn. We calculate $d_y$ for the interfacial Mn in the frozen *f* electrons case, which is 0.70 meV/Mn, very close to $d_y$ in STO (0.73 meV/Mn).

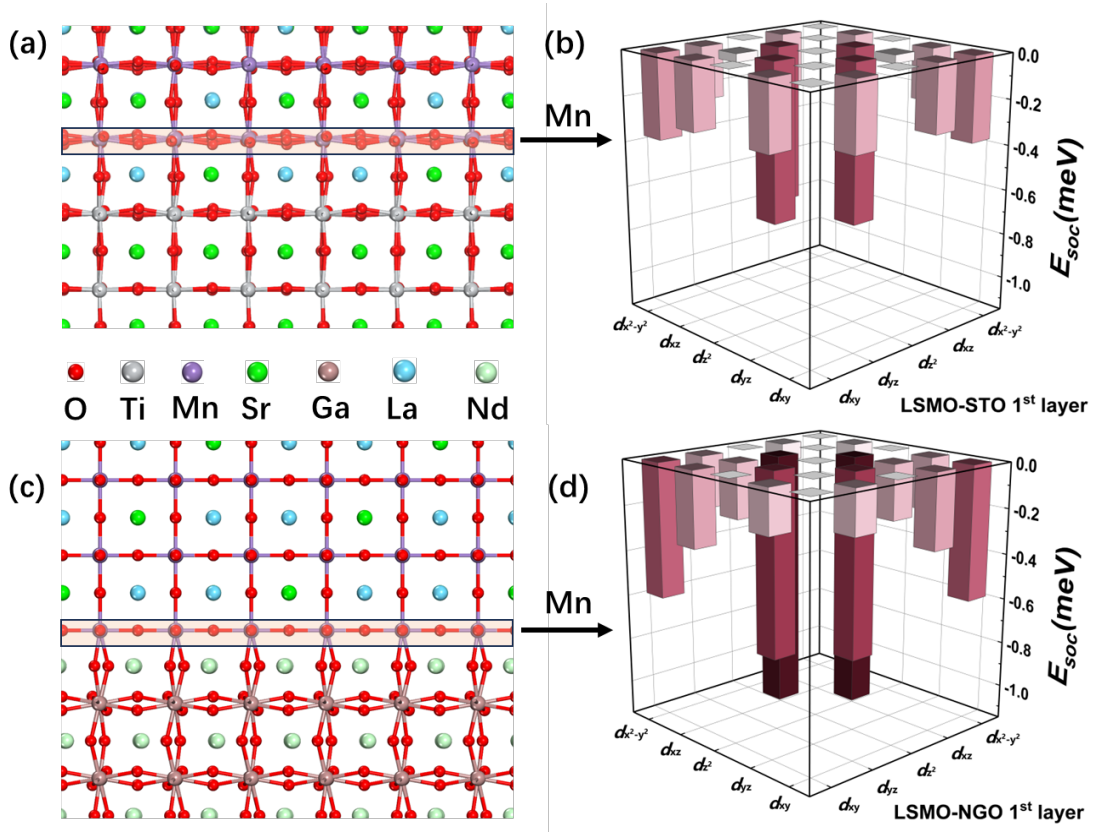

FIG. 4. Calculated atomic structures of (a) LSMO/STO and (c) LSMO/NGO heterostructures. (b) and (d) *d* orbitals resolved SOC matrix elements of the interfacial (1st layer) Mn.

In summary, this study investigates the interfacial DMI in LSMO layers deposited on various substrates. We observed a remarkable range of interfacial DMI constants, spanning from 0.135 to a record-breaking 1.96 pJ/m for magnetic oxide layers. This wide tunability in perovskite LSMO films highlights their potential for applications in chiral spintronics. The variation in DMI is attributed to the breaking of 3*d* orbital symmetry in Mn atoms, which can be controlled through interface engineering. Additionally, the *f*-orbit of Nd atoms in the NGO substrate contributes to the SOC energy. This research offers a novel approach to manipulate and achieve giant DMI, potentially stabilizing topological spin textures in perovskite oxides.

**Acknowledgements:**

The research is supported by the National Natural Science Foundation of China (Nos. 12074025 and 52061135105). W. Wang would like to acknowledge the financial support from National Natural Science Foundation of China (No. 12074071) and National Key Research Program of China (No. 2022YFA1403300). H. Yu would like to acknowledge the support with National Key Research and Development Program of China Grants (No. 2022YFA1402801) and the NSFC under Grants (No.12074026 and 12104208). H. Lu would like to acknowledge the support of National Natural Science Foundation of China (No. 12204027) and Beijing Natural Science Foundation (No. 2232055).

N. Lei, W. Wang, and H. Yu conceived the project and directed the research. L. Yang, X. Zhang, W. Wang and J. Shen are responsible for the thin films' fabrication and X-ray diffractions analysis. H. Wang, J. Wang and H. Yu performed the spin-wave propagation measurements. H. Lu performed the first-principles calculations. L. Yang, Z. Zhao, D. Wei, L. Liu, Y. Yang, D. Pan and J. Zhao conducted the TEM images. L. Yang and N. Lei analyzed the results and wrote the manuscript. All authors contributed to the discussion of the data.